\documentclass{mn2e}
\usepackage{amssymb,amsmath,graphicx,bm,subfigure}

\usepackage[numbers]{natbib}

\bibpunct{(}{)}{;}{a}{,}{,}

\newcommand{\bhat}[1]{\hat{\mathbf{#1}}}
\newcommand{\be}{\begin{equation}}
\newcommand{\ee}{\end{equation}}
\newcommand{\bea}{\begin{eqnarray}}
\newcommand{\eea}{\end{eqnarray}}

\begin{document}

\title{The CMB Derivatives of Planck's Beam Asymmetry}

\author[B. Rathaus and E. D. Kovetz]{Ben Rathaus$^{1}$, Ely D. Kovetz$^{2}$ \\
$^1$Raymond and Beverly Sackler Faculty of Exact Sciences,\\
School of Physics and Astronomy, Tel-Aviv University, Ramat-Aviv, 69978, Israel\\
$^2$Theory Group, Department of Physics and Texas Cosmology Center, \\
The University of Texas at Austin, TX, 78712, USA}

\date{E-mail: ben.rathaus@gmail.com\\ E-mail: elykovetz@gmail.com \\ \\ In original form: \today{}} 
\pagerange{\pageref{firstpage}--\pageref{lastpage}}
\maketitle\label{firstpage}
\begin{abstract}
We investigate the anisotropy in cosmic microwave background Planck maps due to the coupling between its beam asymmetry and uneven scanning strategy. Introducing a pixel space estimator based on the temperature gradients, we find a highly significant ($\sim\!\!20\,\sigma$) preference for these to point along ecliptic latitudes.  We examine the scale dependence, morphology and foreground sensitivity of this anisotropy, as well as the capability of detailed Planck simulations to reproduce the effect, which is crucial for its removal, as we demonstrate in a search for the weak lensing signature of cosmic defects.
\end{abstract}


\section{Introduction}
Cosmic Microwave Background (CMB) satellite experiments from COBE to Planck have suffered from the effects of beam asymmetries, which induce anisotropies in the resulting full sky maps as a result of the uneven scanning strategy of the experiments \citep{Fosalba:2001ey,Chiang:2001mx,Page:2003eu,Wehus:2009zh,Ade:2013dta}. Across the ecliptic plane, the sky pixels are measured mostly in parallel orientations and an ellipticity in the beam response induces anisotropic correlations between them. Meanwhile, these tend to average out near the ecliptic poles, over which multiple passes are made at many different angles. The resulting anisotropy in CMB maps due to this beam-asymmetry-coupled-with-uneven-scanning-strategy (BACUSS) effect has been demonstrated on data from the WMAP experiment using estimators that are sensitive to a quadrupolar modulation on the scales corresponding to the beam sizes of the detectors operating at each frequency, which included a search for $k$-independent violations of rotational symmetry of the primordial power spectrum \citep{Hanson:2009gu, Groeneboom:2009cb, Hanson:2010gu} and a search for a generalized modulation using the Bipolar Spherical Harmonic formalism \citep{Bennett:2010jb,Bennett:2012zja} (see \citet{Kim:2013gka, Ade:2013nlj} for corresponding analyses of Planck data).

In this work we employ a statistical estimator based on temperature gradients \citep{Rathaus}, which was originally designed to detect the weak lensing CMB signature of spherically-symmetric exotic cosmic structures \citep{Itzhaki:2008ih,Fialitzkov}. As we demonstrate below, this estimator proves quite robust for the purpose of studying the BACUSS effect on CMB maps and analysing its sky morphology and dependence on angular scale and astrophysical foregrounds.

\section{Gradient Estimator}
To each direction $\bhat{n}_p$ we assign a score that quantifies to what extent the average temperature gradient tends to be radial with respect to $\bhat{n}_p$ i.e. to point either to or away from the direction $\bhat{n}_p$)
\be\label{eq:estimator}
S_p= \left\langle  (\widehat{\bm{\nabla} T_q}\cdot\bhat{u}_{q,p})^2 \right\rangle_q -0.5,
\ee
where the averaging $\langle \dots \rangle_q$ is over all unmasked sky directions, $\widehat{\bm{\nabla} T_q}$ is the \textit{unit} vector that represents the direction of the temperature gradient at $\bhat{n}_q$, and 
\be
\bhat {u}_{q,p} = \frac{ \bhat{n}_p - \bhat{n}_q(\bhat{n}_p \cdot \bhat{n}_q ) }{1-(\bhat{n}_p \cdot \bhat{n}_q )^2 },
\ee
is the unit vector that lies on the tangent plane to the 2-sphere at $\bhat{n}_q$ and that points in the direction of $\bhat{n}_p$.

In an isotropic universe the direction of the temperature gradient vector at an arbitrary point in the sky is a uniformly distributed random variable. Consider a set of $N$ sky locations such that the angular separation of each pair $q_i,q_j$ of that set is larger than the typical correlation length of CMB temperature gradients. Thus for some $\bhat{n}_p$, the (squared) dot products $(\widehat{\bm{\nabla} T_q}\cdot\bhat{u}_{q,p} )^2$ evaluated at $q_i$ and $q_j$ are uncorrelated, hence they will independently follow a $\cos^2\theta$ probability density function, whose mean is $1/2$ and variance $1/8$. Running the estimator in practice, we transform the map to harmonic space (retaining information up to a desired $\ell_{\rm max}$) and use the HEALPix function \textit{alm2map\_der1} to calculate its gradients. 

\section{Planck Results}
\subsection{Planck SMICA map}
\begin{figure}
\subfigure[~SMICA]{
\includegraphics[width=0.45\columnwidth]{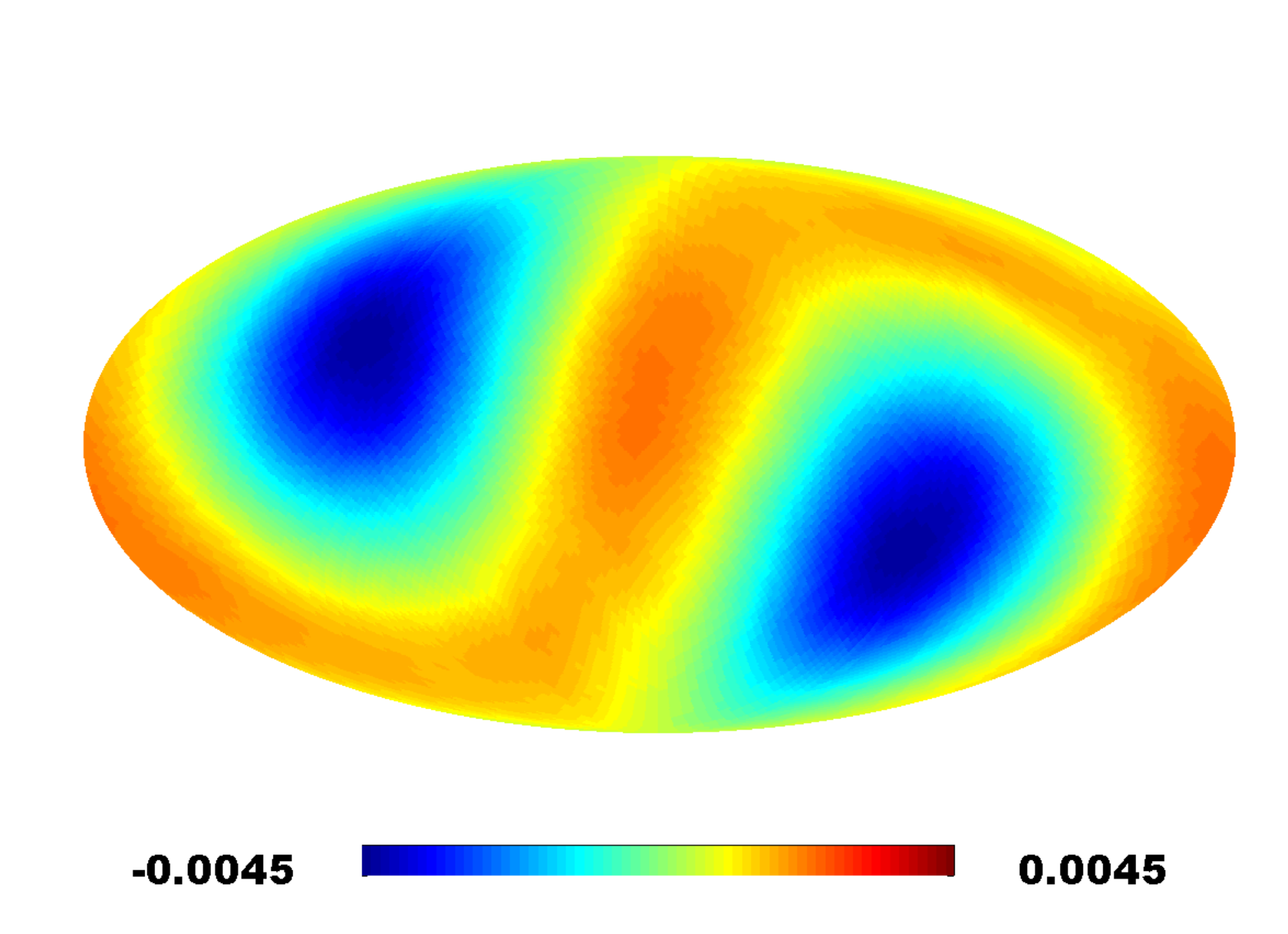}\label{subfig:smica}}
\subfigure[~Random map]{
\includegraphics[width=0.45\columnwidth]{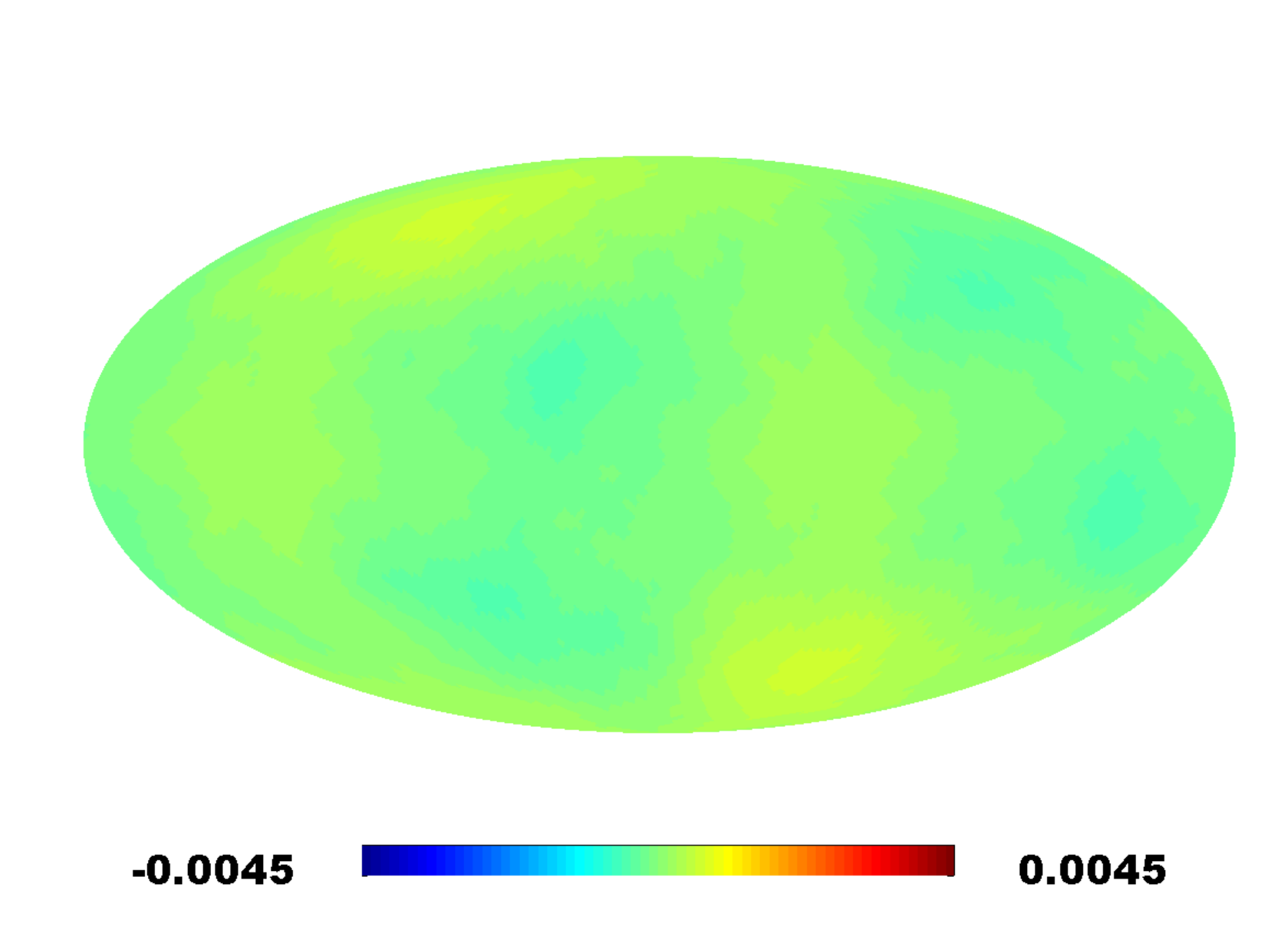}\label{subfig:random}}
\caption{  Gradient score maps for SMICA and a random $\Lambda$CDM realization, with $\ell_{\rm max}=2000$. SMICA confidence mask is applied in both cases.}
\label{fig:RandomAndSMICA}
\end{figure}

The result of the estimator $S_p$ on the Planck SMICA map is shown in Fig.~\ref{fig:RandomAndSMICA}, evaluated at all pixel-centres of a $N_{\rm side}\!=\!32$ HEALPix map. For comparison, we show the result of applying the same estimator to a random $\Lambda$CDM realisation (we use CAMB \citep{Lewis:2002ah} with Planck's best-fit cosmological parameters \citep{Ade:2013zuv} to generate the power spectrum). 
We see that the SMICA score map is highly anisotropic and looks almost like a perfect quadrupole, while the score of the random map is much less pronounced and has no particular shape.

\begin{figure}
\includegraphics[width=0.9\columnwidth]{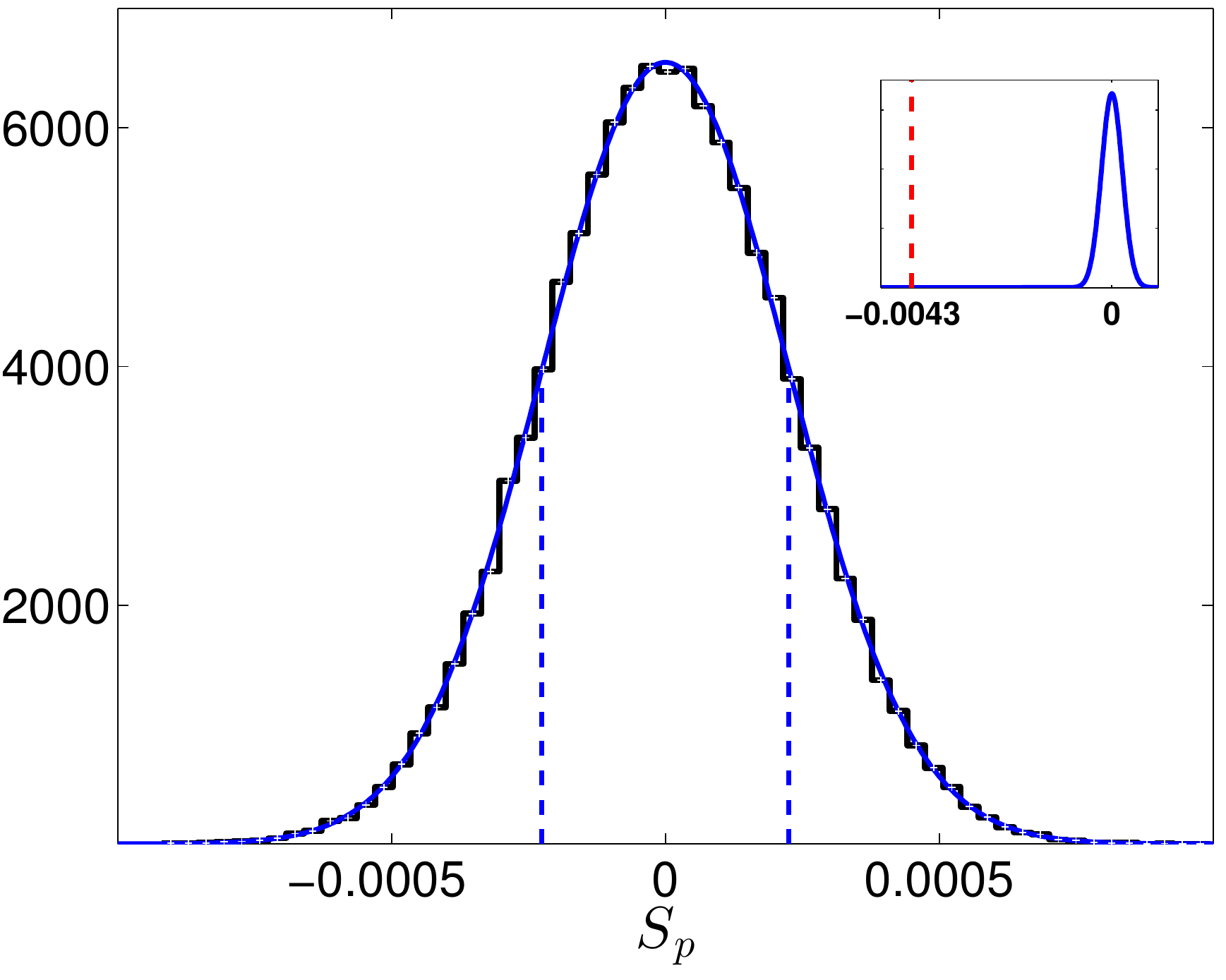}
\caption{  A histogram of single-pixel gradient scores in $10^5$ random $\Lambda$CDM realisations. The SMICA result at the ecliptic pole (red dashed line) lies $\gtrsim19\sigma$ away from $\mu$.}
\label{fig:histogram}
\end{figure}

To estimate the significance of the anisotropy, we generate an ensemble of $S_p$ scores calculated for a randomly selected pixel $p$ in each of $10^5$ random realisations. 
Assuming that $N$ is sufficiently large, the central limit theorem guarantees that we get
\be\label{eq:distribution}
S_p \sim \mathcal{N} \left( \mu=0, \sigma=1/\sqrt{8N} \right).
\ee 
As we show in Fig.~\ref{fig:histogram}, using our set of randomly generated $S_p$, we verify that its distribution follows Eq.~(\ref{eq:distribution}) and using a Gaussian fit we find its standard deviation, thus quantifying the ``effective" number of independent pixels (or the correlation length).
With respect to this distribution, the \textit{local} significance of the minimum value found near the ecliptic pole on the SMICA map lies $19.2\sigma$ away from the expected value of 0. Accounting for the ``look elsewhere" effect (ignoring that the peak shows up at the ecliptic) yields a negligible change to this significance. It is thus evident that the SMICA map exhibits strongly significant anisotropy. 

\subsection{Planck Single-Frequency Maps and Simulations}
A cleaner examination of the BACUSS effect is made possible using a set of single-frequency simulations incorporating the beam properties and scanning pattern of Planck. When applying our estimator to raw single-frequency data, care must be taken to mitigate the effects of foregrounds on the harmonic transform. Therefore, we use the following methodology: we take the 80\% Galactic Planck mask at HEALPix resolution $N_{\rm{side}}=2048$, downsample it to $32$ (corresponding to an angular pixel size $\theta_{pix}=1.8^{\circ}$) and upsample backwards to $2048$ in order to smooth it around the edges, and multiply the input map by the result. Then, after we transform the resulting map to harmonic space and extract its gradients, we extend the 80\% Galactic Planck mask by $\theta_{pix}$ in all directions and combine it with the $5\sigma$ point-source mask of the corresponding frequency and use the result as the mask when calculating the estimator in Eq.~(\ref{eq:estimator}).

\begin{figure}
\subfigure[~$100\,{\rm GHz}$ Simulation]{
	\includegraphics[width=0.45\columnwidth]{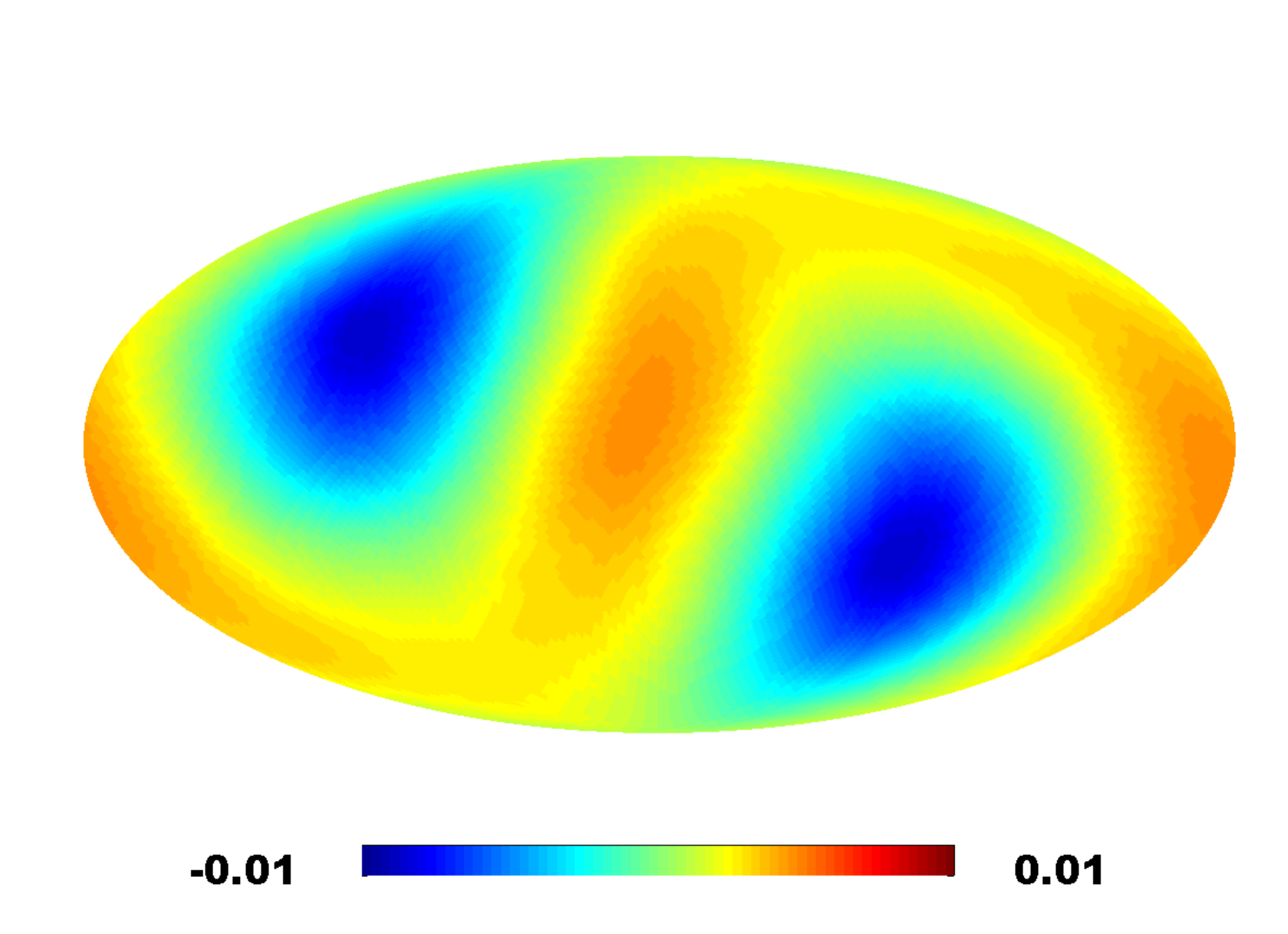}
	\label{subfig:sim100}
}
\subfigure[~$100\,{\rm GHz}$ Real Data]{
	\includegraphics[width=0.45\columnwidth]{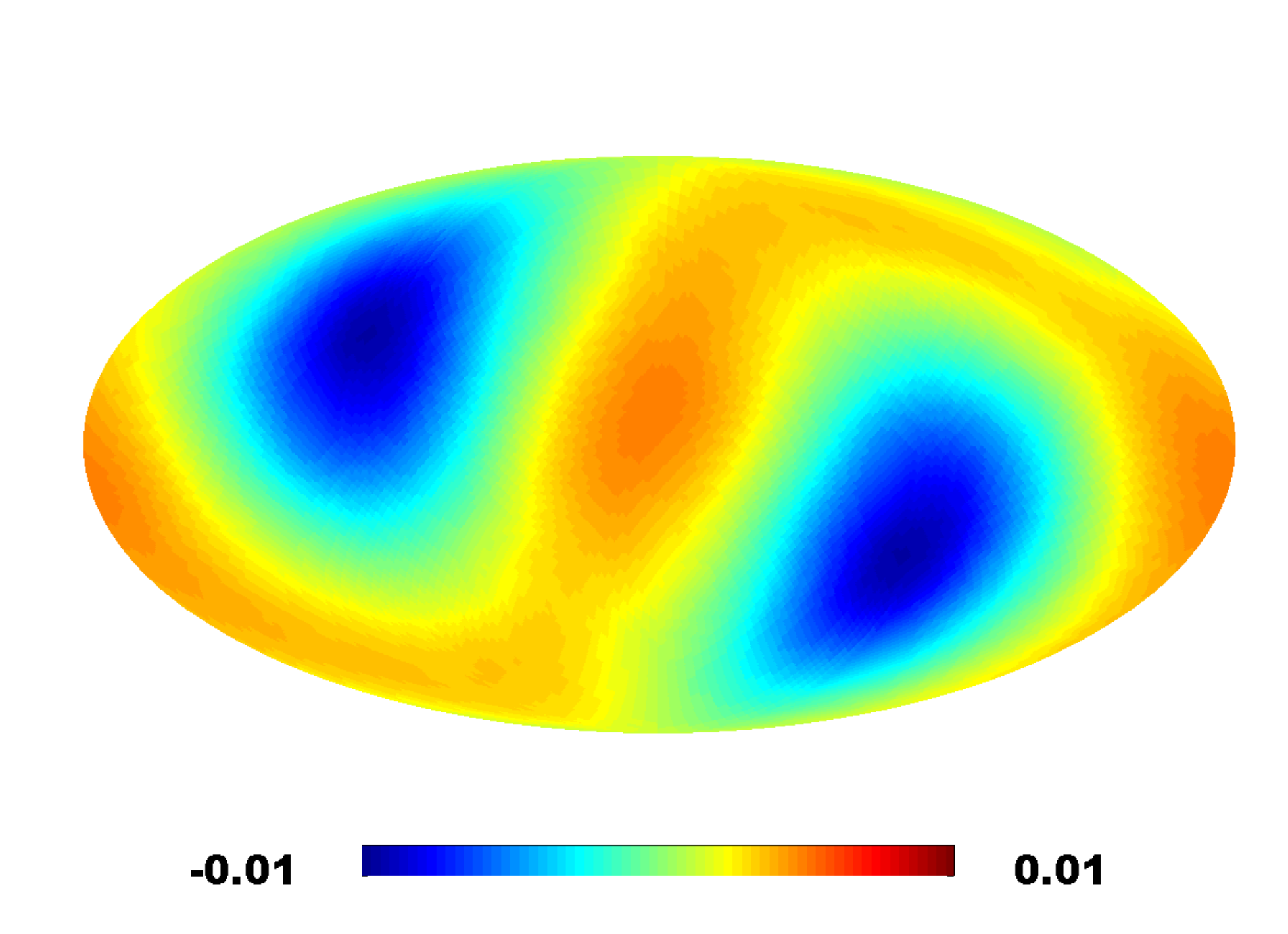}
	\label{subfig:raw100Wu73}
}
\subfigure[~$143\,{\rm GHz}$ Simulation]{
	\includegraphics[width=0.45\columnwidth]{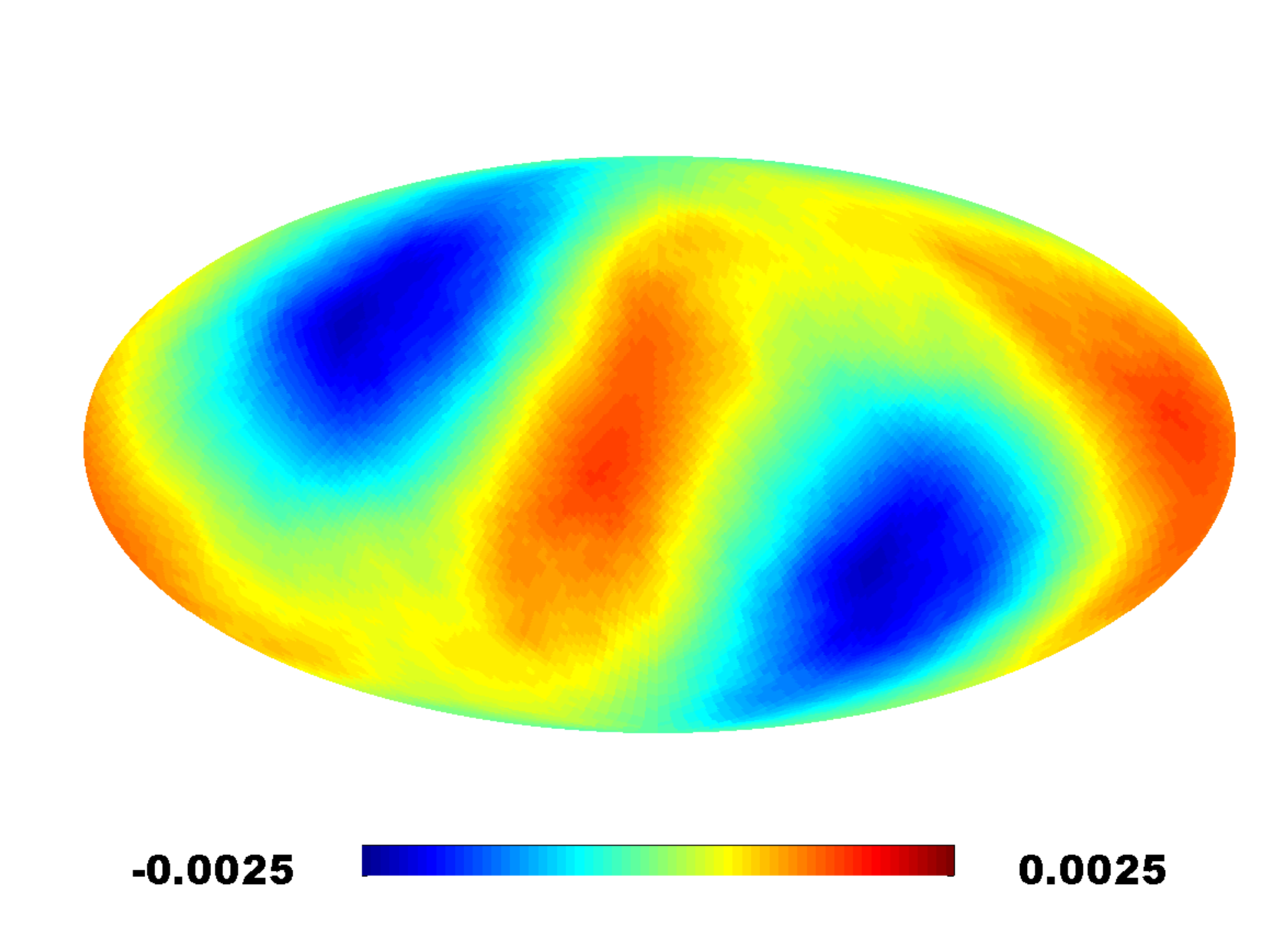}
	\label{subfig:sim143}
} 
\subfigure[~$143\,{\rm GHz}$ Real Data]{
	\includegraphics[width=0.45\columnwidth]{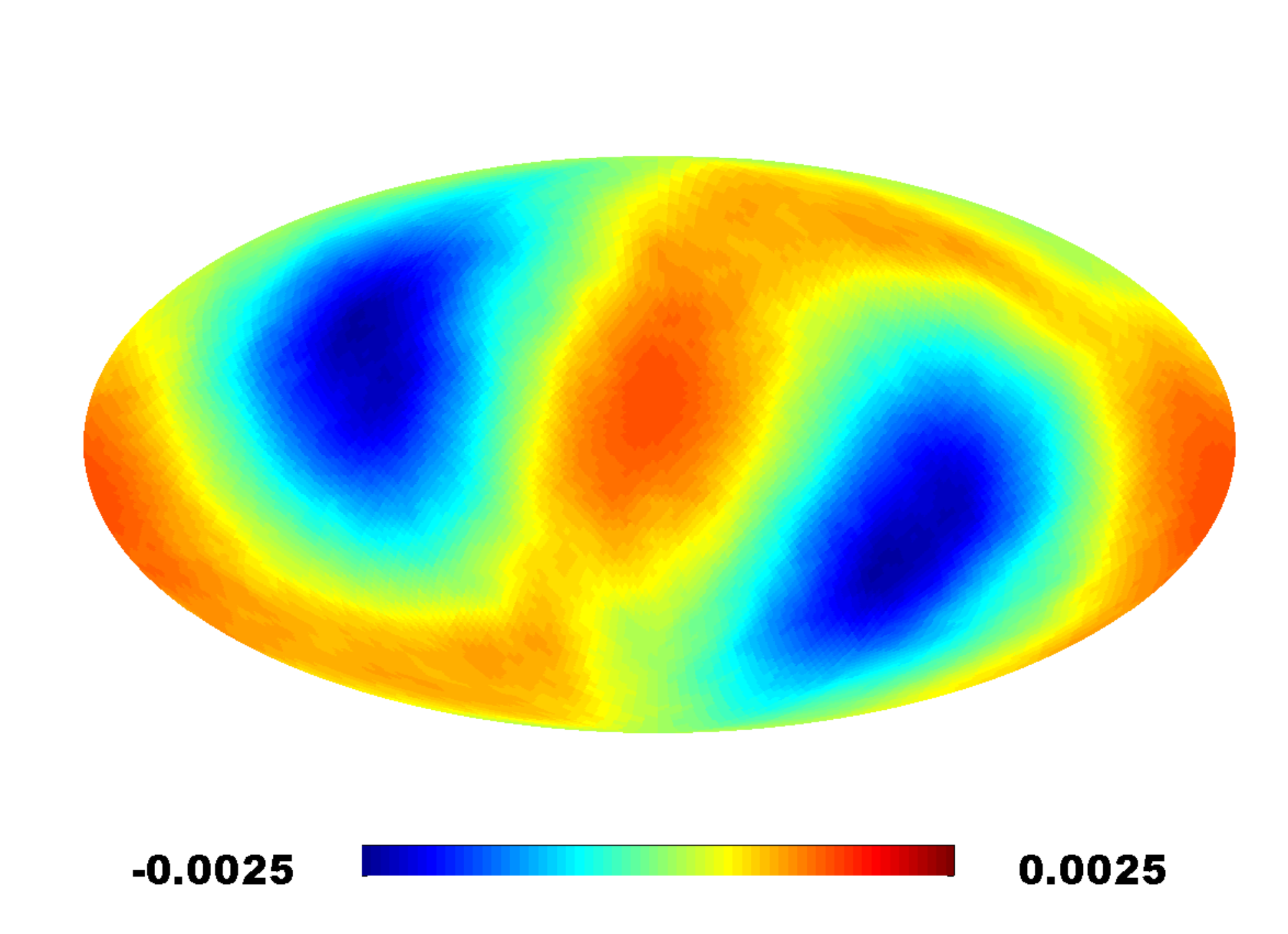}
	\label{subfig:raw143Wu73}
} 
\caption{  Score maps for single frequency \textit{FEBeCOP} simulations and real data, with $\ell_{\max}=1400,1600$ for the $100,143\,{\rm GHz}$ maps, respectively, and masking scheme as described in the text.}
\label{fig:4plot}
\end{figure}

We use two newly released Planck simulations \footnote{\textit{http\,://www.sciops.esa.int/wikiSI/planckpla}} for the $100\,{\rm GHz}$ and $143\,{\rm GHz}$ frequencies of the HFI instrument. The first, generated with the \textit{FEBeCoP} code \citep{Mitra}, uses a pixel space convolution with an effective beam that is calculated at each frequency for each pixel by accumulating the weights of all pixels within a fixed distance from it, summing over all detector observations. In  Fig.~\ref{fig:4plot}, we compare the results of our estimator for these simulations to those for the real data. The second type is generated using the \textit{LevelS} software \citep{Reinecke}, in which for each detector, fiducial time-ordered data are generated and converted to the harmonic domain, where their beam-convolved values are calculated over a three-dimensional grid of sky locations and beam orientations.

\section{Analysing the results}

\subsection{Scale Dependence}
\begin{figure}
	\includegraphics[width=0.85\columnwidth]{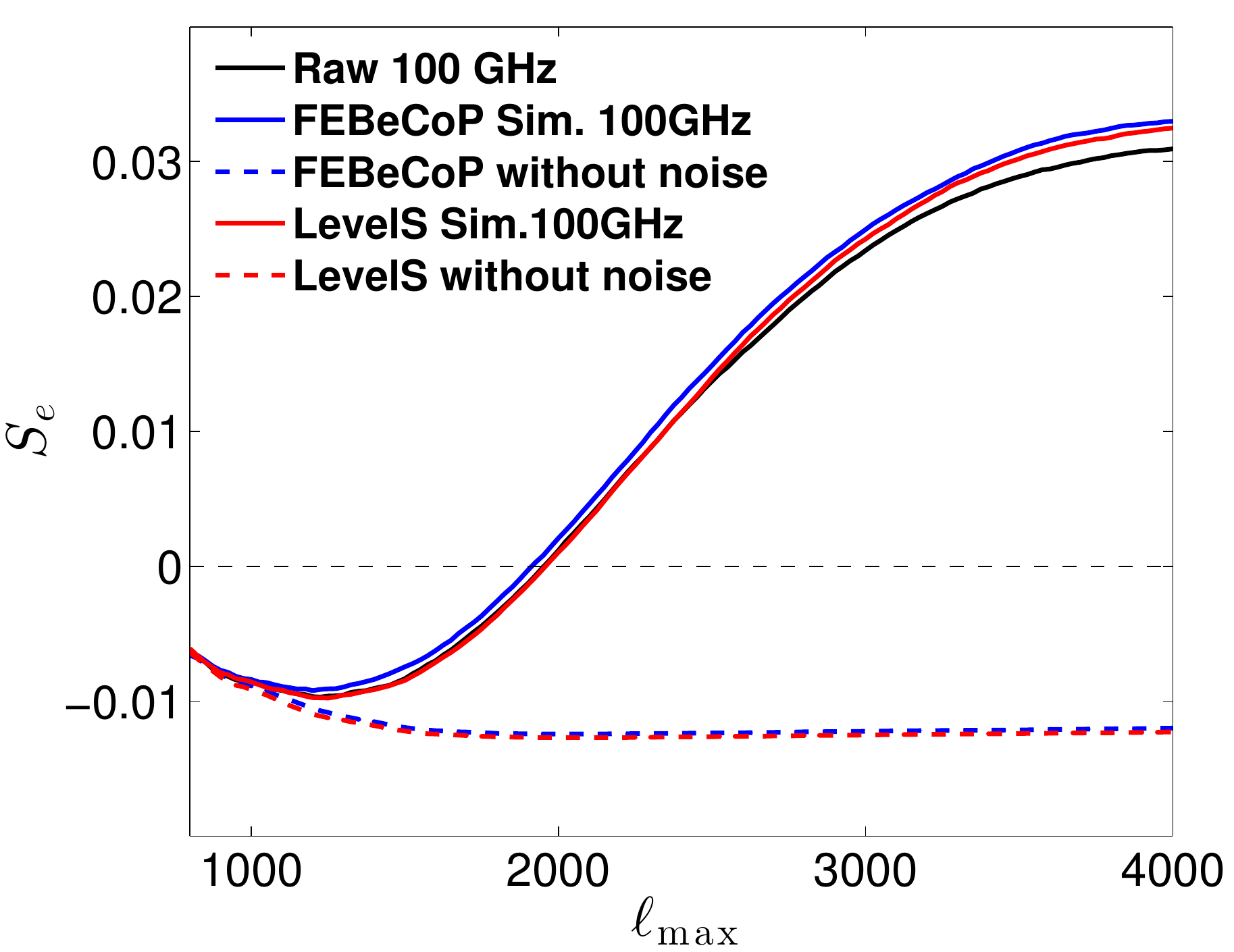}
\caption{ $100\,{\rm GHz}$ maps $S_e$ $\ell_{\max}$-dependence. In noise-added maps (solid), $S_e$ reverses its trend at the beam scale and flips sign further into the noise-dominated regime.}
\label{fig:ellDependence100}
\end{figure}

\begin{figure}
	\includegraphics[width=0.85\columnwidth]{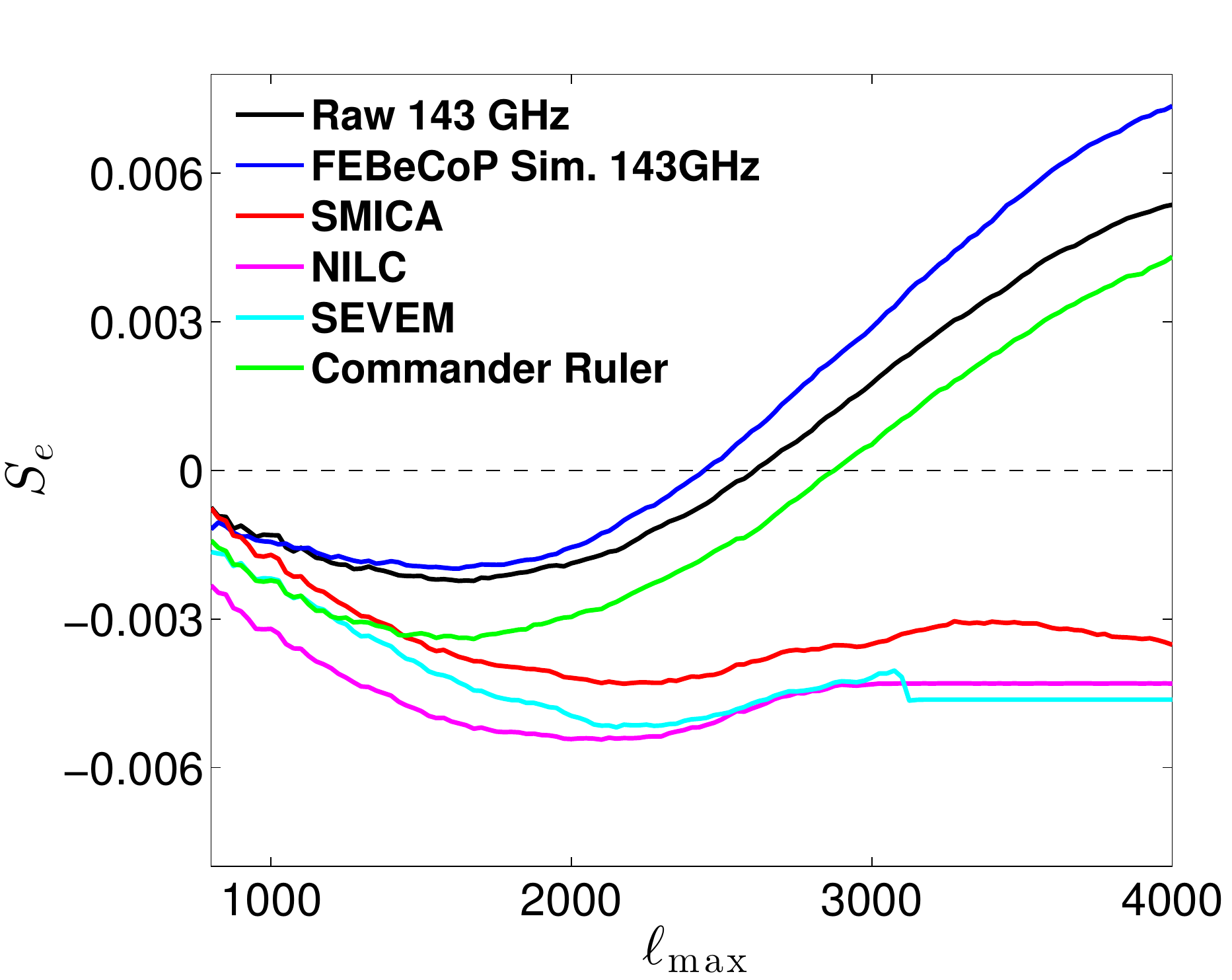}
\caption{  $\ell_{\max}$-dependence of $S_e$ at $143\,{\rm GHz}$ and in the four component separation maps.}
\label{fig:ellDependenceComp}
\end{figure}

In scrutinising the BACUSS effect, we first examine its dependence on the minimal angular scale included in the calculation. In Figs.~\ref{fig:ellDependence100}-\ref{fig:ellDependenceComp} we plot the $\ell_{\max}$-dependence of the BACUSS anisotropy at the ecliptic pole, $S_e$.
Fig.~\ref{fig:ellDependence100} shows the results for the $100\,{\rm GHz}$ frequency, compared to the two types of simulations described above, with and without the instrumental noise contribution. We can see that in the signal-dominated regime, the BACUSS effect induces a strong anisotropy, which peaks (as expected) around the instrumental beam size ($9.65~\rm{arcmin}$), then weakens as we probe into the noise-dominated scales, and diverges as ever-smaller scales are included. This indicates that the correlations induced by the BACUSS effect in the CMB temperature signal are reversed compared to those introduced in the instrumental noise (where the correlation is likely dominated by the 1/f noise contribution \citep{Tegmark:1997vs}).

In Fig.~\ref{fig:ellDependenceComp} we show the corresponding results for the Planck component separation maps, as well as the higher resolution
 ($7.25~\rm{arcmin}$) single frequency map at $143\,{\rm GHz}$. Since we are comparing the component maps to the raw single-frequency maps, in order to remain on the same footing we apply the same masking procedure described above to the component separation maps: smoothing the Galactic component of each confidence mask, zeroing the corresponding map pixels, then extending the Galactic part and combining with the point sources (we have verified that the difference in the result for SMICA, compared to the simpler procedure which was used to generate Fig.~\ref{fig:RandomAndSMICA}, is negligible, but as other component separation maps contain stronger foregrounds residuals, this approach is safer).
The amplitude of the effect in all of these maps is a factor $2\!-\!5$ weaker than in the $100\,{\rm GHz}$ map. The sign-flip of the induced BACUSS anisotropy occurs on smaller scales for the $143\,{\rm GHz}$ maps, as expected given its smaller beam size. We can also see that the component separation maps behave very differently than the single frequency maps, which means that in the absence of tailored simulations which accurately track the component separation method, it is impossible to remove the BACUSS effect from these maps. As we later demonstrate, this can be done to some extent for single frequency maps, at the expense of greater exposure to the effects of residual foregrounds (a more rigorous methodology for overcoming the foreground contribution was used in \citet{Kim:2013gka}, but proved redundant in our case).

\subsection{Sky Morphology}
\begin{figure}
\centering
	\includegraphics[width=0.9\columnwidth]{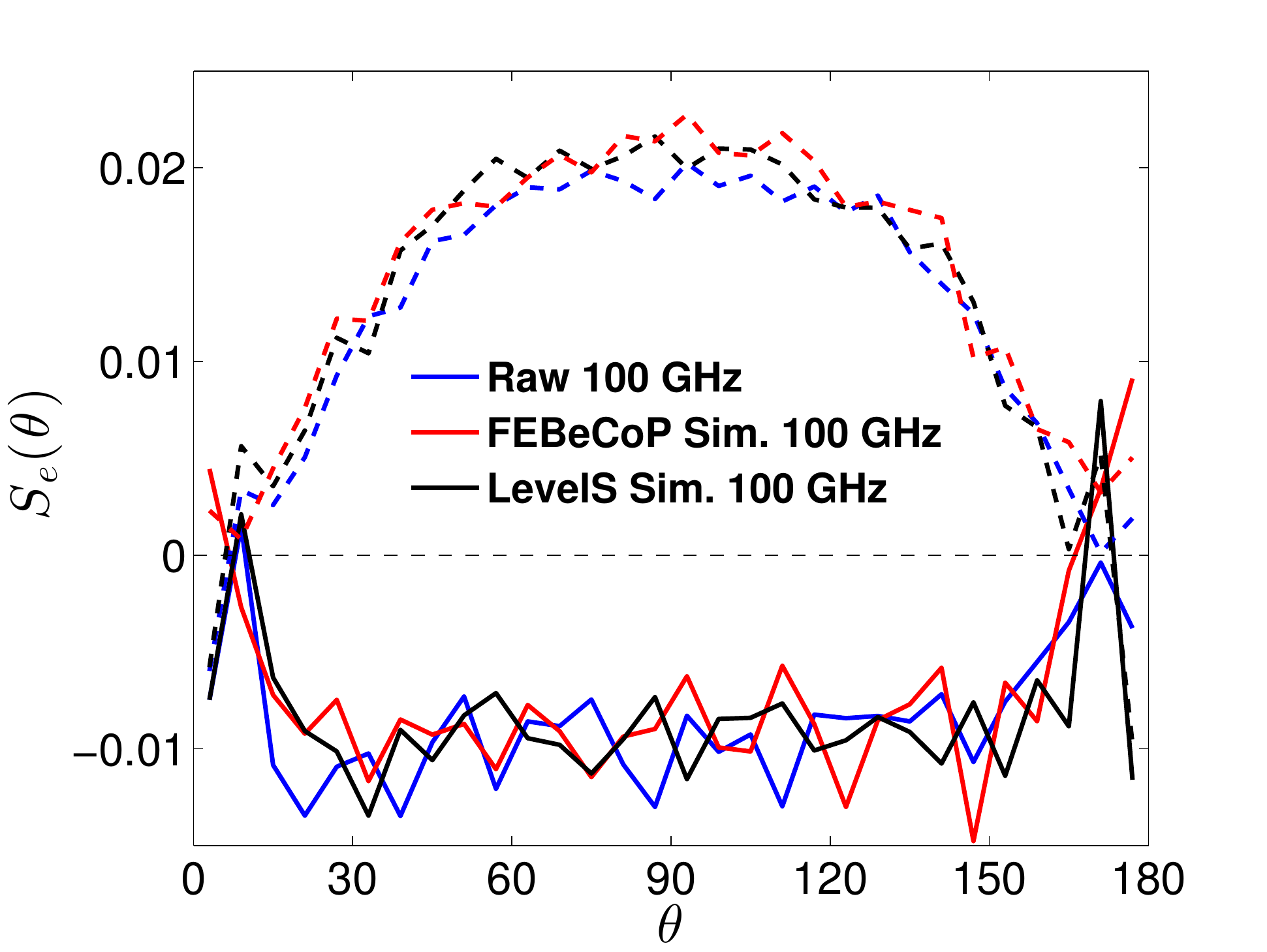}
	\label{subfig:random}
\caption{  Radial profile $S_e(\theta)$ for $100\,{\rm GHz}$ maps, with $\ell_{\rm max}=1400$ (solid) and $\ell_{\rm max}=2600$ (dashed).}
\label{fig:RingDependence}
\end{figure}

It is also interesting to examine the radial profile of the BACUSS anisotropy
\be\label{eq:thetaestimator}
S_e(\theta)= \left\langle  (\widehat{\bm{\nabla} T_q}\cdot\bhat{u}_{q,e})^2 \right\rangle_{\{q(\theta)\}} ,
\ee
where $\theta$ measures the angular distance from the ecliptic pole and $\{q(\theta)\}=\{q,\cos^{-1}(\left|\bhat{n}_q\cdot \bhat{n}_e\right|)=\theta<\pi/30\}$ includes all the unmasked pixels within six degrees of a ring of radius $\theta$ around the ecliptic pole.
In Fig.~\ref{fig:RingDependence}, we plot the radial profile, Eq.~(\ref{eq:thetaestimator}), for the single frequency maps and simulations at $100\,{\rm GHz}$ (corresponding to Fig.~\ref{fig:ellDependence100}). The solid lines were run with $\ell_{\rm max}=1400$ and demonstrate the BACUSS effect on the CMB temperature signal-dominated scales, while the dashed lines were run with $\ell_{\rm max}=2600$ and demonstrate the profile of the induced noise correlations. We see that the effect on the CMB signal w.r.t.\ the ecliptic pole is relatively uniform across most of the map, while the effect on the noise is more highly concentrated around the ecliptic plane. This insight is important for predicting the influence of the BACUSS effect on desired small scale isotropy tests.

\subsection{Foreground Dependence}
\begin{figure}
\centering
\subfigure[~$100\,{\rm GHz}$ Unmasked Data]{
	\includegraphics[width=0.45\columnwidth]{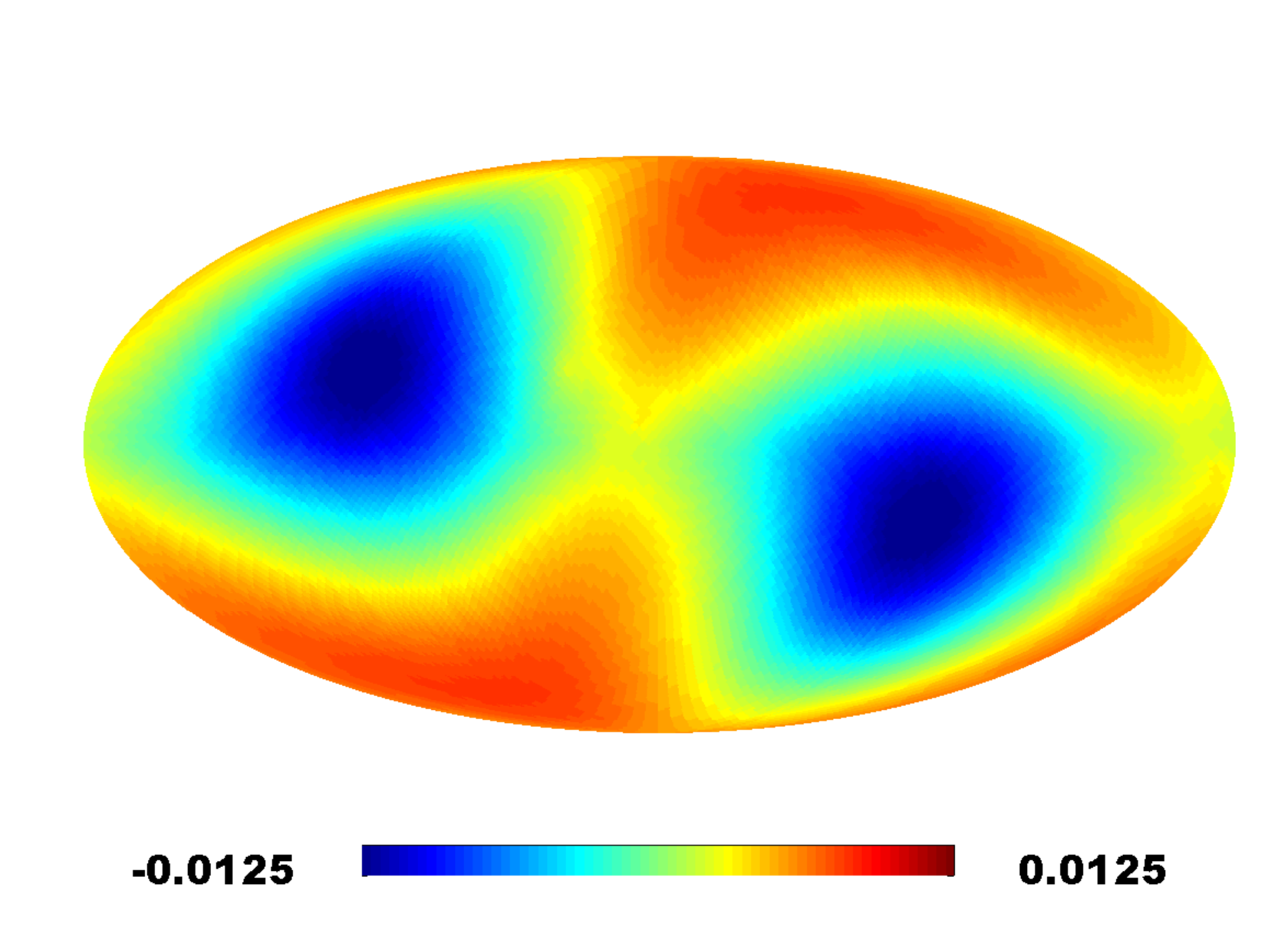}
	\label{subfig:Fore100}
}
\subfigure[~$143\,{\rm GHz}$ Unmasked Data]{
	\includegraphics[width=0.45\columnwidth]{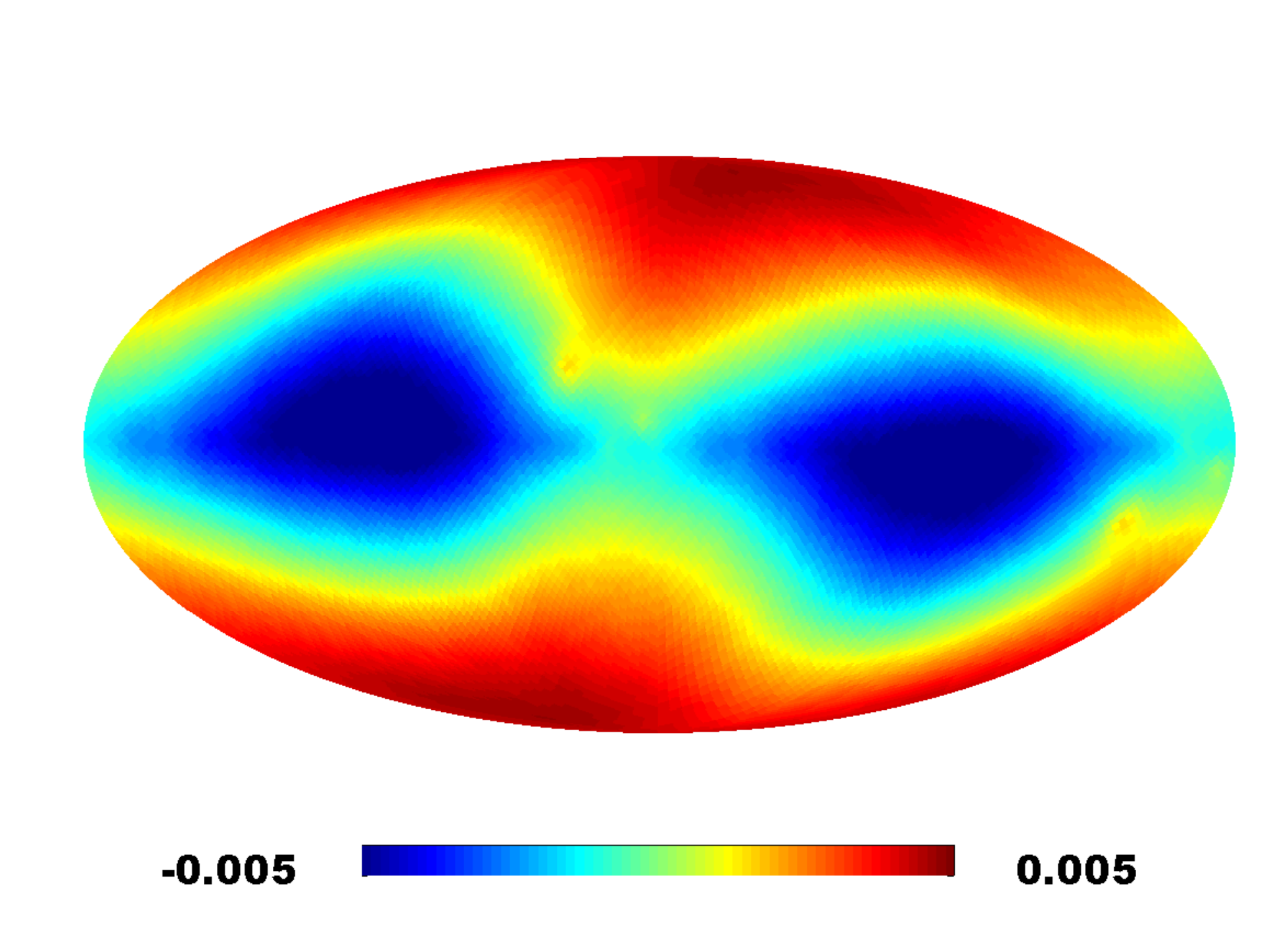}
	\label{subfig:Fore143}
}
\caption{  The score Eq.~(\ref{eq:estimator}) for unmasked single frequency maps at $100\,\rm{GHz}$ and $143\,\rm{GHz}$.}
\label{fig:noMask}
\end{figure}

To gain a better understanding of the influence of foregrounds on the BACUSS effect in Planck maps, we test the behaviour of the gradient score when operating on the raw unmasked data. For this sake we focus again on the $100\,\rm{GHz}$ and $143\,\rm{GHz}$ single-frequency maps. As can be seen in Fig.~\ref{fig:noMask}, while $S_p$ on the unmasked $100\,\rm{GHz}$ map is similar to its masked version (Fig.~\ref{fig:4plot}), the $143\,\rm{GHz}$ result is substantially different. This is due to the fact that the BACUSS effect is much more pronounced in the former, while the foregrounds are stronger in the latter. 

\subsection{Planar Modulation}
One can naively attempt to phenomenologically describe the BACUSS effect as an effective  modulation (with respect to the ecliptic pole) of the intrinsic anisotropy map \citep{Gordon:2005ai}. In this picture, the observed temperature $\tilde T$ at each direction in the sky is related to the true CMB anisotropy at that point by 
\be\label{eq:nabla}
\tilde T(\bhat{n})=\left[  1+ M (\bhat{n}) \right] T(\bhat{n}),
\ee
where $M(\bhat{n})$ is the modulation field. The simplest modulation that can be employed for this purpose is a \textit{quadrupolar} modulation, which is symmetric around the ecliptic plane. In fact, all even-$\ell$ planar modulations share the appealing property of offering an intuitive approximation for the observed BACUSS effect. 

\begin{figure}
\centering
	\includegraphics[width=0.85\columnwidth]{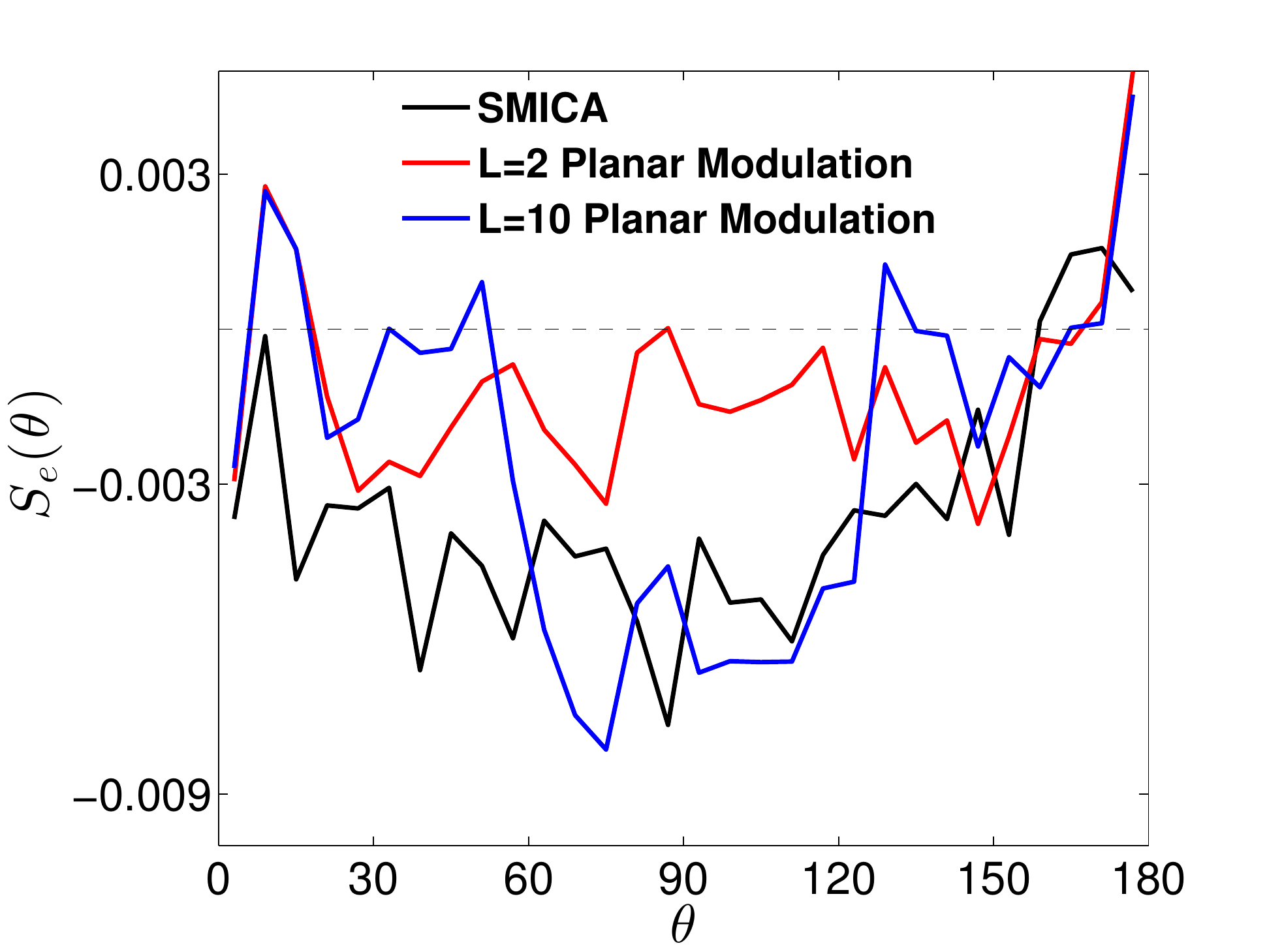}
	\label{subfig:random}
\caption{  Radial profiles $S_e(\theta)$ of the SMICA map and $\ell=2,10$, $\mathcal{A}\!=\!10$ planar modulations of a random CMB realization.}
\label{fig:RingDependenceModulation}
\end{figure}

\begin{figure}
\centering
\subfigure[~Planar $\ell=2$ modulation]{
	\includegraphics[width=0.45\columnwidth]{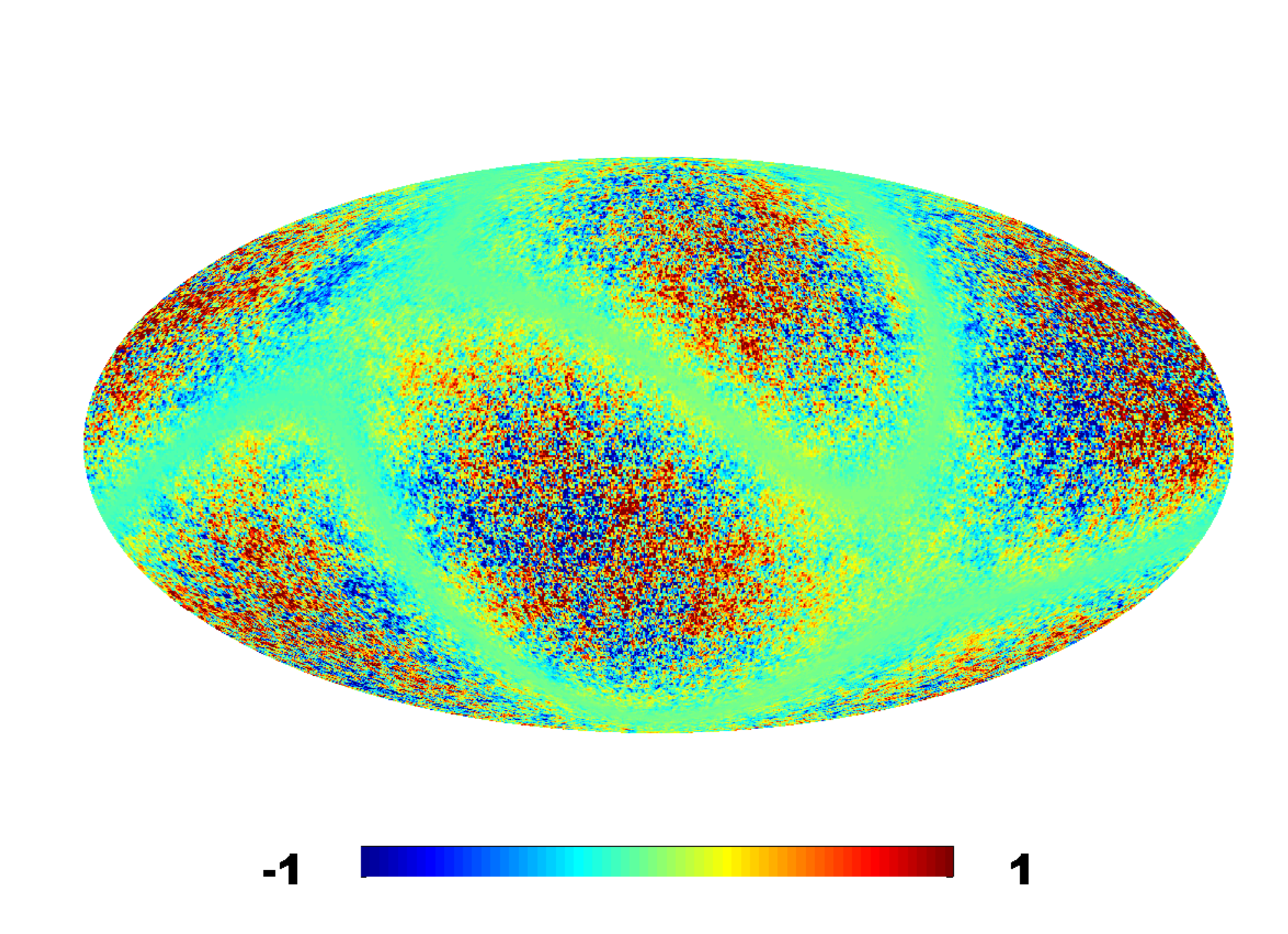}
	\label{subfig:P2}
}
\subfigure[~Planar $\ell=10$ modulation]{
	\includegraphics[width=0.45\columnwidth]{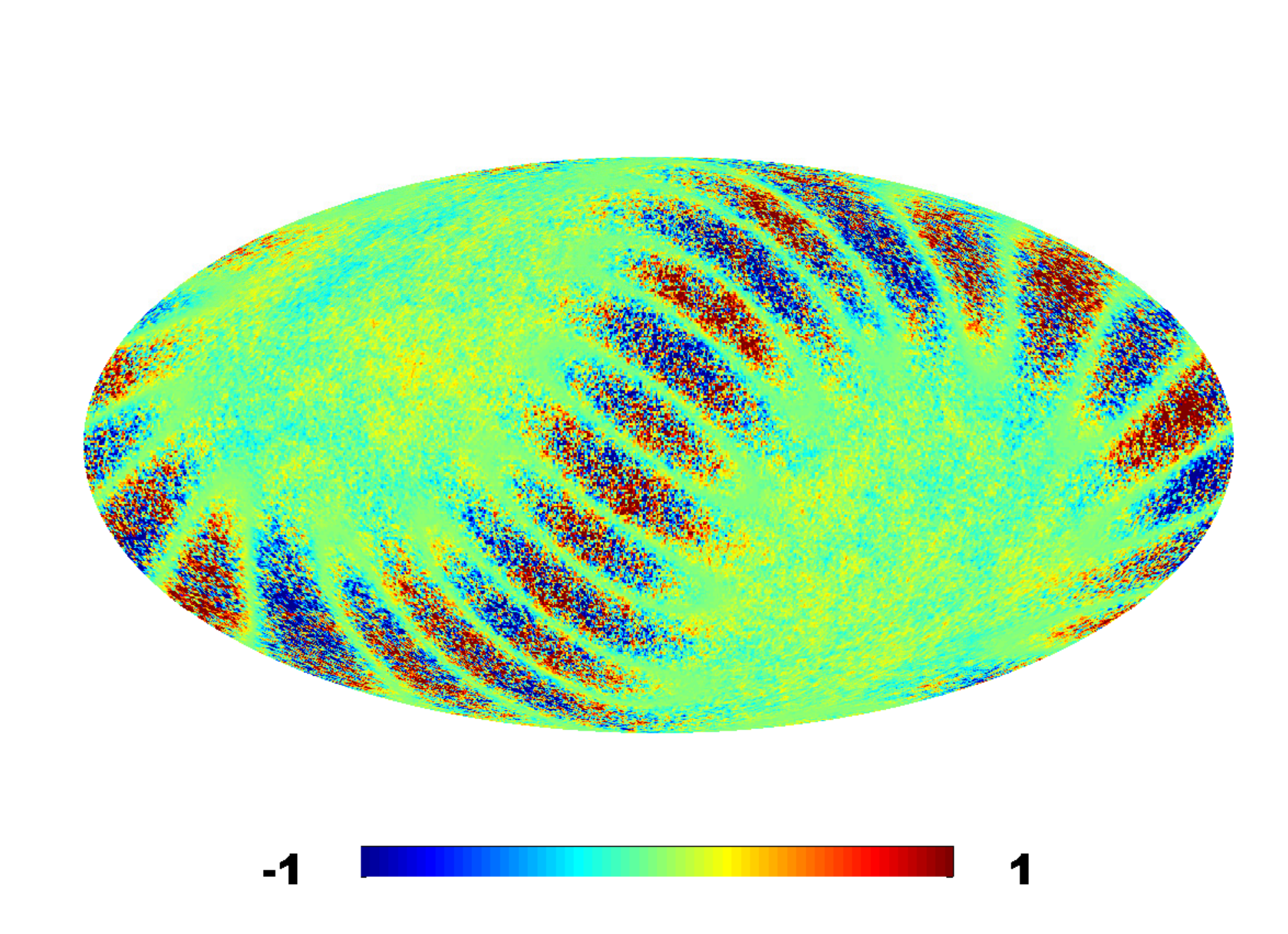}
	\label{subfig:P10}
}
\caption{  $\ell=2,10$, $\mathcal{A}\!=\!10$ planar modulations of a random CMB realization (in units of $\mu K$).}
\label{fig:ModulatedMaps}
\end{figure}

Let us consider a general even-$\ell$ \textit{planar} modulation, by setting $m=\ell$. The modulation field is then 
\be\label{eq:mod}
M(\bhat{n}) = \mathcal{A}\left(Y_{\ell,\ell}(\bhat{n}) +(-1)^\ell Y_{\ell,-\ell}(\bhat{n})\right)\propto \sin^\ell \theta \cos \ell\phi
\ee
where $\mathcal A$ is the amplitude of the modulation and $\theta,\phi$ are the usual spherical angles, with respect to the modulation axis (the following arguments are equally valid for $M\propto \sin^\ell \theta \sin \ell\phi$).
The gradient at each point in the modulated map is
\be\label{eq:modGrad}
\bm\nabla \tilde T(\bhat{n})  =   \left[ 1+M(\bhat{n})\right] \bm\nabla T(\bhat{n})+ T(\bhat{n})  \bm\nabla M(\bhat{n}).
\ee
While the first term in Eq.~(\ref{eq:modGrad}) points in a random direction, the derivative in the second term yields:
\bea
\bm\nabla M(\bhat{n}) &=& \bhat{e}_\theta \partial_\theta M + \bhat{e}_\phi(\partial_\phi M) /\sin\theta\nonumber\\
&=&\bhat{e}_\theta \ell (\sin^{\ell-2} \theta\sin 2\theta \cos \ell\phi)/2  \nonumber\\
&&-\ell \bhat{e}_\phi \sin^{\ell-1}\theta \sin \ell\phi
\eea
Around $\theta=90^\circ$ the first term, which points in the $\bhat{e}_\theta$ direction, becomes very small. For a given $\phi$, however, the amplitude of the second term is maximal. Therefore, for such a modulation the tangential component of the gradient, is amplified around $\theta=90^\circ$. 
Since with respect to each direction $\bhat{n}_p$ there are more pixels in the vicinity of the plane perpendicular to it, the most prominent contribution to the overall score comes from that portion of the map, and the modulation in Eq.~(\ref{eq:mod}) will thus yield a minimum score at the ecliptic poles, as we see in Planck data.
However, a closer look at specific examples of modulated maps shows that a planar modulation is a poorly effective model for the BACUSS effect. We consider planar modulations of a randomly-generated realisation of $\Lambda$CDM with $\ell=2,10$, where the amplitude had to be set to $\mathcal{A}\!=\!10\,$(!)\! so that the overall gradient score $S_p$ of the modulated maps will be similar to that of the SMICA map.
As shown in Fig~\ref{fig:RingDependenceModulation}, the radial profile of the modulated maps is substantially different from that of the SMICA map. 
In addition, it is apparent from Fig.~\ref{fig:ModulatedMaps}, which shows the modulated maps in these two cases, that in order to reproduce the SMICA result, the modulation amplitude must be extremely large, yielding an overwhelmingly anisotropic map.

\section{Test Case:\ A Search for Cosmic Defects}
As described in \citet{BeNaSunny}, an anomalously large structure whose gravitational potential stretches over cosmological distances will induce an observable signal on the CMB via its lensing potential. A particular example for such a lens is   
a spherically-symmetric overdense structure induced by a pre-inflationary particle (PIP) which remains within the visible universe after inflation \citep{Itzhaki:2008ih,Fialitzkov} (other examples include the lensing signature of a cosmic texture or a large void, see \citet{Das:2008es,Masina,Kovetz:2012jq} and references within).
In \citet{Rathaus}, a similar estimator to $S_p$ in Eq.~(\ref{eq:estimator}), limited to a band surrounding the pixel $p$ (instead of summing over the whole map), was proposed in order to detect the weak lensing signature of a PIP.
As mentioned above, when performing such a search for small scale anisotropy in the CMB, the BACUSS effect must first be efficiently removed from the map. To demonstrate the compensation for the BACUSS effect using detailed simulations which incorporate the beam asymmetry and scanning pattern, we performed the search suggested in \citet{Rathaus} by subtracting the average of $30$ \textit{FEBeCOP} simulations at $143\,{\rm GHz}$ from the real data (both maps taken up to $\ell_{\rm max}=1600$), applying the estimator Eq.~(\ref{eq:estimator}) within the radii $\theta=5^\circ-60^\circ$ around each point in the sky and using the single-frequency masking procedure described above. No significant evidence for lensing by a cosmic defect was found, placing a (mild) constraint on the existence of such defects in general and ruling out the specific PIP scenario at the focus of \citet{Rathaus}.

\section{Discussion}
Focusing on the temperature gradients, we have shown that Planck maps exhibit significant anisotropy due to the BACUSS effect.  We analysed the scale dependence of this anisotropy and found that the effect is reversed in the signal- and noise-dominated regimes. We demonstrated that the effect is maximised at the scale corresponding to the instrumental beam size, and that its amplitude differs by as much as a factor of $5$ between frequencies. We used the radial profile of the induced anisotropy to examine the uniformity of the effect across the map and used it to disqualify naive phenomenological descriptions of it such as quadrupolar planar modulations. Our method also proved to be weakly sensitive to foregrounds, as the use of foreground masking is straightforward and efficient in pixel space.

As Planck simulations incorporating the asymmetric beam shape and the actual scanning patterns were shown here to trace the real single-frequency data quite accurately, the prospects for removing the effect efficiently look promising. We have demonstrated this for the example of searching for the weak lensing effect of large cosmic defects (without an understanding of the BACUSS effect and its removal, constraining such models would not have been possible). 
Nevertheless, it should be kept in mind that this removal is valid only for estimators such as $S_p$, and is not necessarily possible for any small scale analysis of Planck CMB maps in general. 
Furthermore, we emphasise that care should be taken with component separation maps, e.g. SMICA (which has been recommended for preforming isotropy tests of the CMB), as detailed pipeline simulations for these maps have not been made publicly available. 
Finally, we point out that these anisotropies could very well persist in Planck's polarisation maps as well (we have verified that a BACUSS effect is seen in WMAP polarisation maps at $94\,{\rm GHz}$). Unless carefully mitigated, this may affect the analysis of the lensing contribution to B-modes at small scales \citep{SPTPol,Polarbear}. It will be interesting to explore this in the near future.

\vspace{-0.1in}
\section*{Acknowledgments}
We thank Tom Crawford, Sunny Itzhaki and Eiichiro Komatsu for very useful discussions. We acknowledge the use of the HEALPix package \citep{healpix} and the Planck Legacy Archive \footnote{\textit{http\,://pla.esac.esa.int/pla/aio/planckProducts.html}} (PLA). EDK was supported by the National Science Foundation under Grant Number PHY-1316033.

\label{lastpage}

\end{document}